\documentclass[usenatbib,useAMS]{mnras}

\usepackage{graphicx}	
\usepackage{amsmath}	
\usepackage{amssymb}	
\usepackage{multicol}        
\usepackage{bm}		
\usepackage{pdflscape}	

\def \aap{A\&A}
\def \apjl{ApJ}

\def \mnras{MNRAS}

\def \aaps{A\&AS}
\def \aj{AJ}

\def \apjs{ApJS}

\def \araa{ARA\&A}
\bibpunct{(}{)}{;}{a}{}{,}

\newcommand{\Teff}{T_\mathrm{eff}}
\newcommand{\logg}{\log g}
\newcommand{\FeH}{\mathrm{[Fe/H]}}

\newcommand{\kms}{km~s$^{-1}$}

\usepackage[T1]{fontenc}
\usepackage{ae,aecompl}

\usepackage{newtxtext,newtxmath}

\title{The GALAH Survey: Accurate Radial Velocities and Library of Observed Stellar Template Spectra}

\author[T. Zwitter et al.]{
Toma\v{z} Zwitter,$^{1}$\thanks{Contact e-mail: \href{mailto:tomaz.zwitter@fmf.uni-lj.si}{tomaz.zwitter@fmf.uni-lj.si}}
Janez Kos,$^{2}$ 
Andrea Chiavassa,$^{3}$ 
Sven Buder,$^{4}$ 
Gregor Traven,$^{1}$\newauthor
Klemen \v{C}otar,$^{1}$ 
Jane Lin,$^{5}$
Martin Asplund,$^{5,6}$ 
Joss Bland-Hawthorn,$^{2,6}$ 
Andrew~R. Casey,$^{7,8}$\newauthor 
Gayandhi De Silva,$^{2,9}$ 
Ly Duong,$^{5}$
Kenneth C.\ Freeman,$^{5}$ 
Karin Lind,$^{4,10}$
Sarah Martell,$^{11}$\newauthor     
Valentina D'Orazi,$^{12,13,7}$ 
Katharine J. Schlesinger,$^{5}$ 
Jeffrey~D. Simpson,$^{9,11}$
Sanjib Sharma,$^{2,6}$\newauthor
Daniel~B. Zucker,$^{13,9}$ 
Borja Anguiano$^{14,13}$ 
Luca Casagrande,$^{5,6}$ 
Remo Collet,$^{15}$\newauthor 
Jonathan Horner,$^{16}$ 
Michael J. Ireland,$^{5}$
Prajwal R. Kafle,$^{17}$ 
Geraint Lewis,$^{2}$\newauthor 
Ulisse Munari,$^{12}$  
David M.\ Nataf,$^{18}$ 
Melissa Ness,$^{4}$
Thomas Nordlander,$^{5,6}$\newauthor 
Dennis Stello,$^{2,6,11,15}$ 
Yuan-Sen Ting,$^{19,20,21}$   
Chris G.\ Tinney,$^{11}$
Fred Watson,$^{9,22}$\newauthor 
Rob A. Wittenmyer,$^{16}$ 
Maru\v{s}a \v{Z}erjal$^{5}$
\\
(Affiliations listed after the references)
}

\date{Last updated 2018 April 16; in original form 2018 March 31}

\pubyear{2018}

\begin{document}
\label{firstpage}
\pagerange{\pageref{firstpage}--\pageref{lastpage}}
\maketitle

\begin{abstract}
GALAH is a large-scale magnitude-limited southern stellar spectroscopic survey. Its second data release (GALAH DR2) provides values of stellar parameters and abundances of 23 elements for 342,682 stars (Buder et al.). Here we add a description of the public release of radial velocities with a typical accuracy of 0.1~\kms\ for 
336,215 of these stars,  
achievable due to the large wavelength coverage, high resolving power and good signal to noise ratio of the observed spectra, but also because convective motions in stellar atmosphere and gravitational redshift from the star to the observer are taken into account. 
In the process we derive medians of observed spectra which are nearly noiseless, as they are obtained from between 100 and 1116 observed spectra belonging to the same bin with a width of 50~K in temperature, 0.2~dex in gravity, and 0.1~dex in metallicity. Publicly released 1181 median spectra have a resolving power of 28,000  and trace the well-populated stellar types with metallicities between $-0.6$ and $+0.3$. 
Note that radial velocities from GALAH are an excellent match to the accuracy of velocity components 
along the sky plane derived by Gaia for the same stars. 
The level of accuracy achieved here is adequate for studies of dynamics {\it within\/ } stellar clusters, 
associations and streams in the Galaxy. So it may be relevant for studies of the distribution of dark matter. 
\end{abstract}






\section{Introduction}

At first sight, measurement of stellar radial velocity (RV) seems a rather trivial task. This view seemed to be reflected in a vivid debate on the reorganisation of IAU commissions three years ago, when an opinion was advanced that this is a ``mission accomplished''. Indeed, measurement of stellar RVs has a long tradition, dating back to the 19$^\mathrm{th}$ century (\citealt{vogel1873,seabroke1879,seabroke1887,seabroke1889}), with perhaps the first large RV catalogue being the one of \citet{wilson53}. A huge success in detection of exoplanets using modern spectrographs with excellent RV measurement precision \citep[e.g.\ ][]{pepe04} supports the same impression. On the other hand this century has seen the first large scale stellar spectroscopic surveys, increasing the number of observed objects from 14,139 stars of the Geneva-Copenhagen survey \citep{nordstrom04}, which observed one star at a time, to hundreds of thousands observed with modern multiple fibre spectrographs, as part of the RAVE (\citealt{steinmetz06,kunder17}), Gaia-ESO \citep{gilmore12}, APOGEE (\citealt{holtzman15,majewski16}, LAMOST \citep{liu17}, and GALAH (\citealt{deSilva15,martell17}) surveys. These efforts have been focused on Galactic archaeology \citep{freeman02}, which aims to decipher the structure and formation of our Galaxy as one of the typical galaxies in the universe through detailed measurements of stellar kinematics and chemistry of their atmospheres. RV measurement is needed to derive the velocity vector of a star and so permit calculation of action variables of its motion in the Galaxy \citep[e.g.\ ][]{mcmillan08}. Given a large span of Galactic velocities and an internal velocity dispersion of $\gtrsim 10$~\kms\ seen in different Galactic components, an accuracy of $\sim 1$~\kms\ in RV, which is generally achieved by the mentioned spectroscopic surveys, seems entirely adequate. Such accuracy can be obtained by a simple correlation of an observed spectrum with a well matching synthetic spectral template, thus supporting the simplistic view of RV measurements we mentioned.  

But we are on the brink of the next step in the measurement of the structure and kinematics of the Galaxy. The European Space Agency's mission Gaia \citep{prusti16} is about to publish its second data release (Gaia DR2), which is expected to provide parallaxes at a level of $\sim 40 \mu$as and proper motion measurements at a level of $\sim 60 \mu$as yr$^{-1}$ for a Solar type or a red clump star at a distance of 1~kpc (we use equations from \citet{prusti16} and neglect any interstellar extinction). This corresponds to a distance error of 4\%\ and an error in velocity along the sky plane of 0.3~\kms. So it is desirable that also the RV is measured at a similar level of accuracy. A special but very important case are stars in stellar clusters, associations, and streams. Most of these aggregates have such a small spread in distance that even Gaia cannot spatially resolve them, but Gaia can clearly and unambiguously identify stars which are members of any such formation. Note also that such clusters are generally loosely bound, with an escape velocity at a level of a few \kms. Knowledge of the velocity vector at a level of $\sim 0.1$~\kms\ therefore allows us to study internal dynamics of stellar aggregates. In a special case of spherical symmetry it also permits us to judge the position of an object within a cluster. Stars with a large radial velocity and small proper motion vs.\ the cluster barycentre are generally at a distance similar to the cluster centre. But those that have a  small radial velocity component, a large proper motion, and projected location on the cluster core are likely to be in front or at the back end of the cluster. 

Deriving RVs of stars at a 0.1~\kms\ level is a challenge. Note that we refer to accuracy here, while exoplanet work, for example,  focuses on precision. Also, we want to measure the RV component of the star's barycentre. So effects of convective motions in the stellar atmosphere and of gravitational redshift need to be addressed. Here we attempt to do so for data from the GALAH survey. The paper is structured as follows: in the next three sections we briefly discuss the observational data, data reduction pipeline, and properties of our sample. In Section 5 we present details of the RV measurement pipeline. The results are discussed in Section 6, data products in Section 7, and Section 8 contains the final remarks. 

\section{Observational data}

GALAH (The Galactic Archaeology with HERMES) is an ambitious stellar spectroscopic survey observing with the 3.9-m Anglo-Australian Telescope of the Australian Astronomical Observatory (AAO) at Siding Spring. The telescope's primary focal plane, which covers $\pi$ square degrees, is used to feed light from up to 392 simultaneously observed stars into the custom designed fibre-fed HERMES spectrograph with four arms, which cover the wavelength ranges of 4713--4903~\AA\ (blue arm), 5648--5873~\AA\ (green arm), 6478--6737~\AA\ (red arm), and 7585--7887~\AA\ (IR arm).  The fibres have a diameter of 2~arc~sec on the sky. Details of the instrument are summarised in \citet{sheinis15}. The resulting spectra have a resolving power of $R = 28,000$. A typical signal to noise ratio (S$/$N) per resolution element for survey targets is 100 in the green arm. This is achieved after three 20-min exposures for targets with $V \lesssim 14.0$; for bright fields the exposure time is halved. In case of bad seeing the total exposure time is extended using additional exposures within the same night. 'A spectrum' in this paper refers to data from a given star collected during an uninterrupted sequence of scientific exposures of its field. Its effective time of observation is assumed to be mid-time of the sequence.  

GALAH is a magnitude limited survey of stars with $V < 14.0$, where the $V$ magnitude is computed from 2MASS magnitudes as discussed in \citet{martell17}. Stars are selected randomly with no preference for colour or other properties. To avoid excessive reddening or crowding most of the targets are located at least 10 degrees from the Galactic plane. The upper limit on the distance from the plane is set by the requirement that there should be $\sim 400$ stars with $V<14.0$ in the observed field of view with a diameter of $2^\mathrm{o}$. 

\section{Data reduction pipeline}

The GALAH data are reduced with a custom-built pipeline, which is fully described in \citet{kos17}; here we use results of its version 5.3. The pipeline produces wavelength-calibrated spectra with effects of flat-fielding, cosmetic effects, fibre efficiency, optical aberrations, scattered light, fibre cross-talk, background subtraction, and telluric absorptions all taken into account. Wavelength is calibrated using spectra from a ThXe arc lamp, which are obtained before or after each set of science exposures. The pipeline output includes an initial estimate of radial velocity (RV\_synt), which is determined through  correlation of a normalised observed spectrum shifted to a Solar barycentric frame with a limited number of synthetic template spectra. This radial velocity has a typical uncertainty of 0.5~\kms, which is accurate enough to allow alignment of observed spectra at their approximate rest wavelengths, as the RV uncertainty is much smaller than the resolution of the spectra.

Our radial velocity measurement routines require estimates of the following values of basic stellar parameters: the effective temperature ($\Teff$), surface gravity ($\log g$), and metallicity, which in this case actually means iron abundance ($\FeH$). Values of these labels are estimated by a multi-step process based on SME (spectroscopy made easy) routines, which are used as a learning set for {\it The Cannon} data-driven approach, as described in \citet{buder18}. 

\section{The sample}

\begin{figure}
  \includegraphics[width=0.38\columnwidth,angle=270]{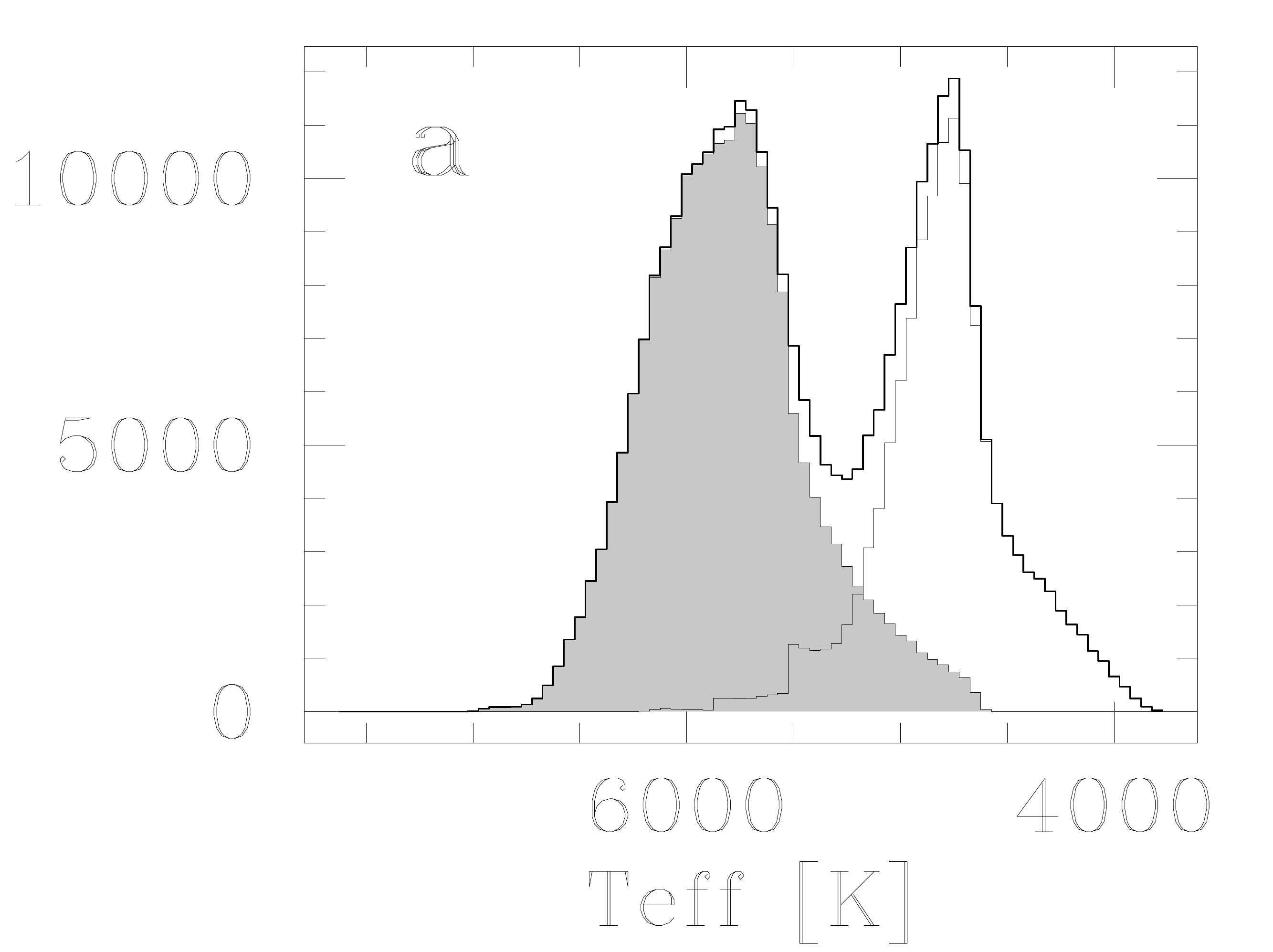}                   
  \includegraphics[width=0.38\columnwidth,angle=270]{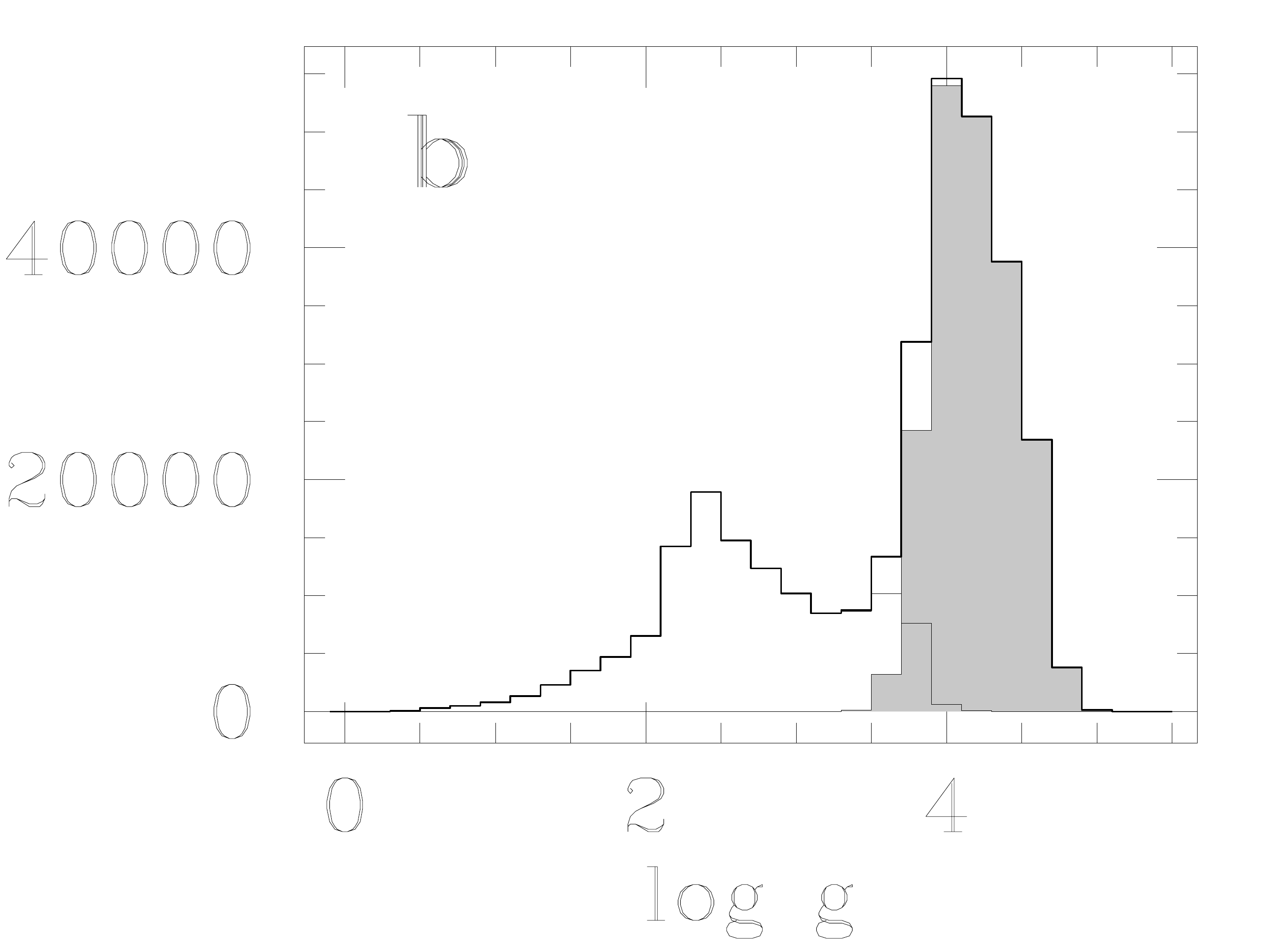}
  \includegraphics[width=0.38\columnwidth,angle=270]{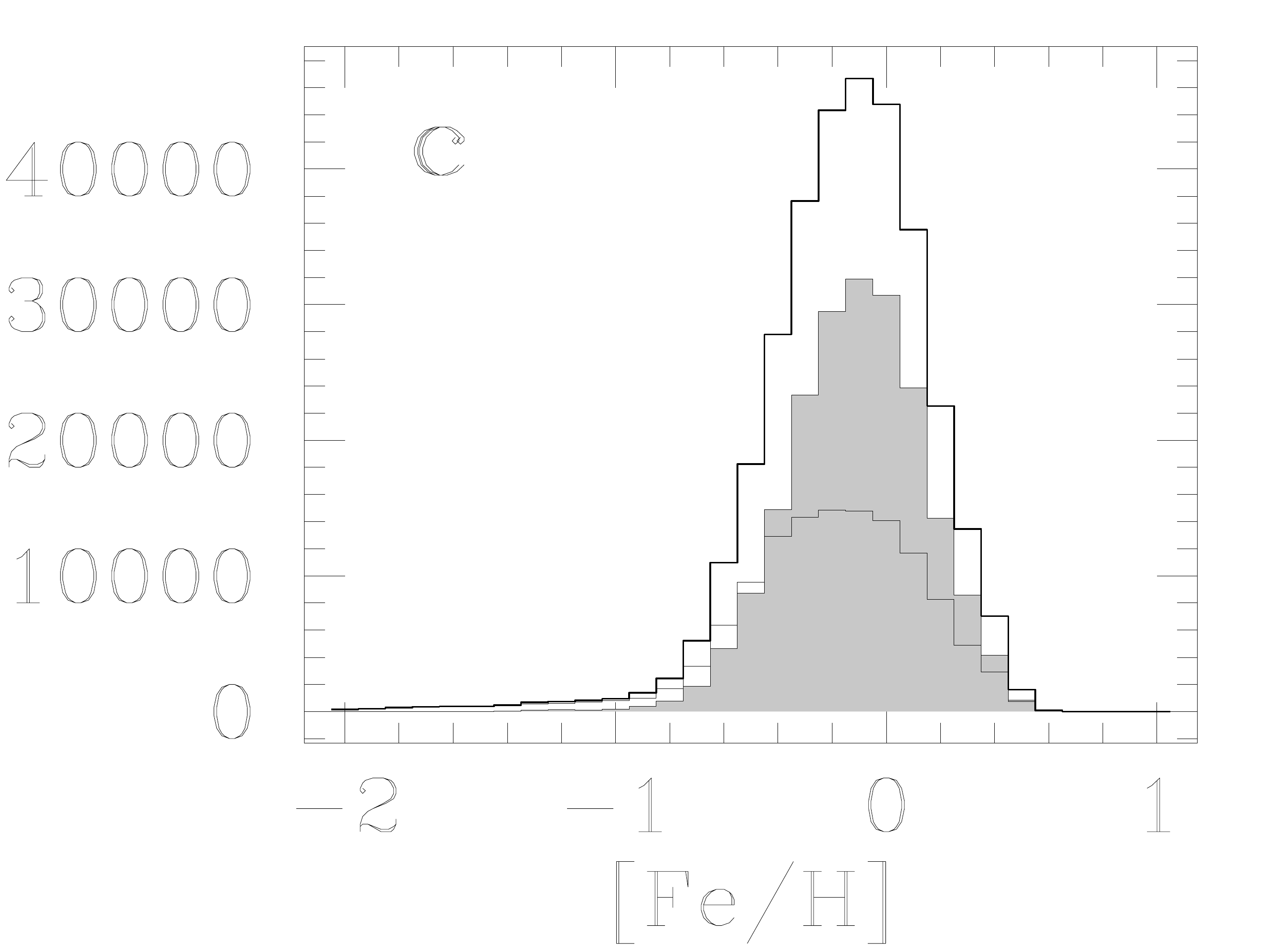}
  \includegraphics[width=0.38\columnwidth,angle=270]{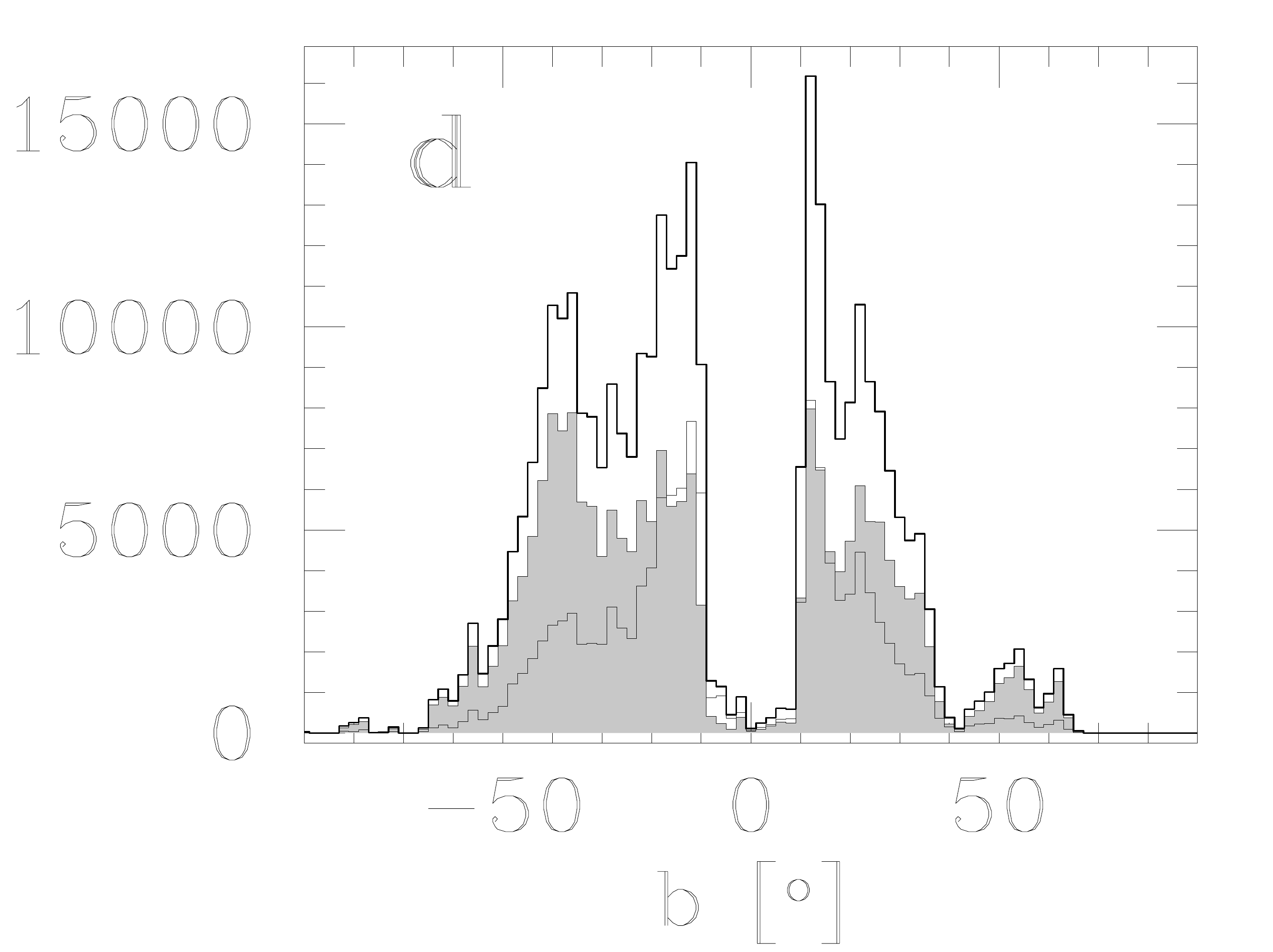}
  \caption{Distribution of spectra for which we computed radial velocities as a function of temperature (a), surface gravity (b), metallicity (c), and Galactic latitude (d). The three histograms are for dwarfs (shaded grey), giants, and their sum. The line dividing these two classes is defined in Eq.\ 1.
  }
  \label{Figparameterhistos}
\end{figure}

In this paper we publish RVs for 336,215 objects out of a total of 342,682 which form the second data release of the GALAH survey (GALAH DR2). The difference between the two numbers are observations with bookkeeping issues on exact time of observation which prevent accurate computation of heliocentric corrections. Some of the published velocities may be influenced by systematic problems which are identified by three kinds of flags: (1) the reduction flag points to possible problems encountered during observations, (2) the synt flag to problems in initial estimate of radial velocity, and (3) the parameter flag to problems with parameter determination during {\it The Cannon} data driven approach, frequently due to peculiarities like binarity or presence of emission lines. These flags are summarised in the value of {\sc flag\_cannon} (see \citet{buder18}, Table 5). GALAH DR2, which is based on {\it The Cannon} analysis with an internal release date of 8th January 2018, contains 213,231 spectra without such problems, i.e.\ with all warning flags set to 0. Here we publish RVs for 212,734 of these objects. Altogether our database contains 514,406 RVs (from these 323,949 with all flags set to 0) derived with the same approach. Additional spectra are either repeated observations of the same objects or objects observed as part of other programs than GALAH  which will be published separately. We plan to add radial velocities for more spectra in the future, which will include new observations but also hot and cool stars, for which no Cannon-derived values of stellar parameters are available at this time. Figure~\ref{Figparameterhistos} shows distributions in temperature, gravity, metallicity, and Galactic latitude for stars with our RVs. We distinguish between giants and dwarfs, with the dividing line assumed to be 
\begin{equation}
\log g = g_1 + (g_2-g_1)(T_1-\Teff)/(T_1-T_2)
\end{equation}
where $g_1 = 3.2$, $g_2 = 4.7$, $T_1 = 6500$~K, and $T_2 = 4100$~K. Note that the fraction of giants in Figure~\ref{Figparameterhistos}d falls off as we move away from the Galactic plane: an un-reddened red clump star with an apparent magnitude $V=13.5$ is at a distance of $\sim 4$~kpc, which, unless observed close to the Galactic plane, puts it well into the sparsely populated halo.

\section{Measurement of radial velocities}

Here we review details of the RV measurement process. We discuss (i) the input data, (ii) construction of a reference library of medians of observed spectra, (iii) a library of synthetic spectra computed from 3-dimensional models, which allow for convective motions in stellar atmospheres, (iv) computation of radial velocity shifts of the observed spectrum vs. the observed median spectrum, (v) of the observed median spectrum vs. synthetic model, and, finally, (vi) the allowance for gravitational shift of light as it leaves the stellar surface to start its travel towards the Earth. In summary, measurement of RVs consists of the following main steps: (a) we compute median observed spectra and derive RV of each star vs. median spectra, (b) we derive zero-point for median spectra by comparing with a synthetic library which allows for 3-dimensional convective motions, (c) we further improve RV by accounting for gravitational redshift.

\subsection{Input data and construction of observed median spectrum library}

Input from the data reduction pipeline supplies an un-normalised spectrum of the observed object with subtracted background and telluric absorptions. It is calibrated in a wavelength frame tied to the AAO observatory. The spectrum represents light collected during an uninterrupted sequence of scientific exposures of a given object, which normally lasts about an hour. The reference time at mid-exposure is used to translate this spectrum into the heliocentric reference frame using the \textsc{Iraf} routine {\em rvcorr} in its {\em astutil} package. We use the J2000.0 coordinates of the source as input, any proper motion is neglected. The observed spectrum is also normalised via the {\em continuum} non-interactive routine in \textsc{Iraf}, using 3 pieces of cubic spline for each of the four arms, with a symmetric rejection criterion of $\pm 3.5 \sigma$, 10 rejection iterations and a growing radius of 1 pixel. This normalisation is not intended to provide a classical normalised spectrum, which would have a value of 1.0 in the continuum regions away from spectral lines, because classical normalisation requires careful adaptations depending on the object's type. What we want is to make spectra of stars of similar types compatible with each other. A low order continuum fit with symmetric rejections is entirely adequate for this purpose. 

Next, we adopt the values of stellar parameters for the spectrum determined by {\it The Cannon} routines, as described  by \citet{buder18}. Values are rounded to the nearest step in the $N \Delta T$ ladder in temperature, to $N \Delta \log g$ in gravity, and to $N \Delta \FeH$\ in metallicity, where $N$ is an integer, $\Delta T = 50$~K, $\Delta \log g = 0.2$~dex, and   
$\Delta \FeH = 0.1$dex. These rounded values now serve as labels that indicate to which stellar parameter bin our spectrum belongs. Even if the uncertainties in stellar parameters for each observed spectrum might be somewhat larger than our bins, we statistically beat them down by doing an averaging in each bin. Systematic uncertainties of stellar parameter values are typically much smaller than bin size. Note that stellar rotation is generally slow in our sample and, similarly to other line broadening mechanisms, it is very similar for spectra within a given spectral parameter bin.

Finally, we use the initial value of radial velocity as determined by the data reduction pipeline and which is reported in the value of the column RV\_synt in \citet{buder18}. As described in \citet{kos17}, these velocities are determined by a cross correlation of the observed spectrum in the blue, green, or red arm (normalised with a single spline) with a set of 15  AMBRE  model  spectra \citep{delaverny12}, which span 4000 -- 7500  K  in  temperature,  with  250  K  intervals, while values $\logg = 4.5$ and $\FeH = 0.0$ are assumed throughout. The value of radial velocity is determined as an average of values determined from the blue, green, and red arms, unless any of the values deviate by more than 3~\kms, in which case it is omitted from the average. These radial velocities have typical errors of $\sim 0.5$~\kms, which is good enough to shift the observed spectrum to an approximate rest wavelength.
\label{Sec51} 

\begin{figure*}
\includegraphics[width=1.11\columnwidth,height=1.8\columnwidth,angle=270]{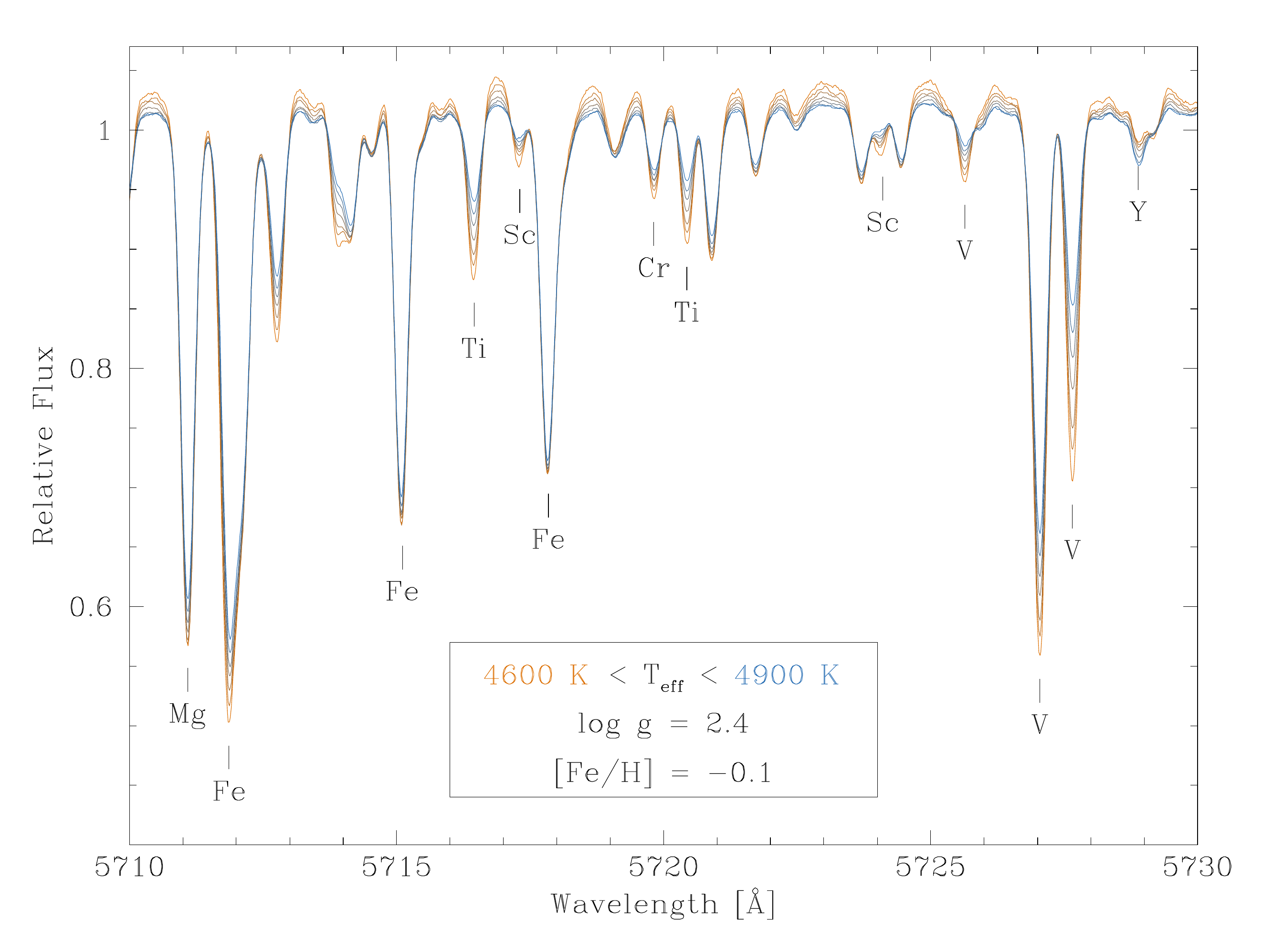}                   %
  \caption{A sequence of median observed spectra in the green arm. Their Cannon-determined labels differ in temperature  (in steps of 50~K), as indicated.}
  \label{Figexamplemedians}
\end{figure*}

All observed spectra are first resampled to logarithmic wavelength steps of $\Delta \lambda / \lambda = 300,000^{-1}$. The median observed spectrum in a given spectral parameter bin is then calculated as a median of all observed shifted spectra in that bin, where only spectra with {\it The Cannon} parameter flag set to 0 are considered. This condition excludes a vast majority of spectra with any type of peculiarity, such as binarity or presence of emission lines, while other properties, such as stellar rotation, tend to be very similar within a given spectral bin. Median flux is calculated for each wavelength step separately. Using median instead of average is useful to avoid influence of any outliers with spectral peculiarities. Combining many nearly identical spectra makes the observed median spectrum virtually noise-free compared to any individual spectrum. Figure \ref{Figexamplemedians} shows examples of median spectra for a small wavelength span within the green channel. 

Note that there is no guarantee that the median spectrum has exactly zero radial velocity. Also use of medians on bins with a small number of contributing spectra can be problematic. These issues are addressed next. 

\subsection{Calculation of radial velocity}

\begin{figure}
  \includegraphics[width=0.68\columnwidth,height=1.0\columnwidth,angle=270]{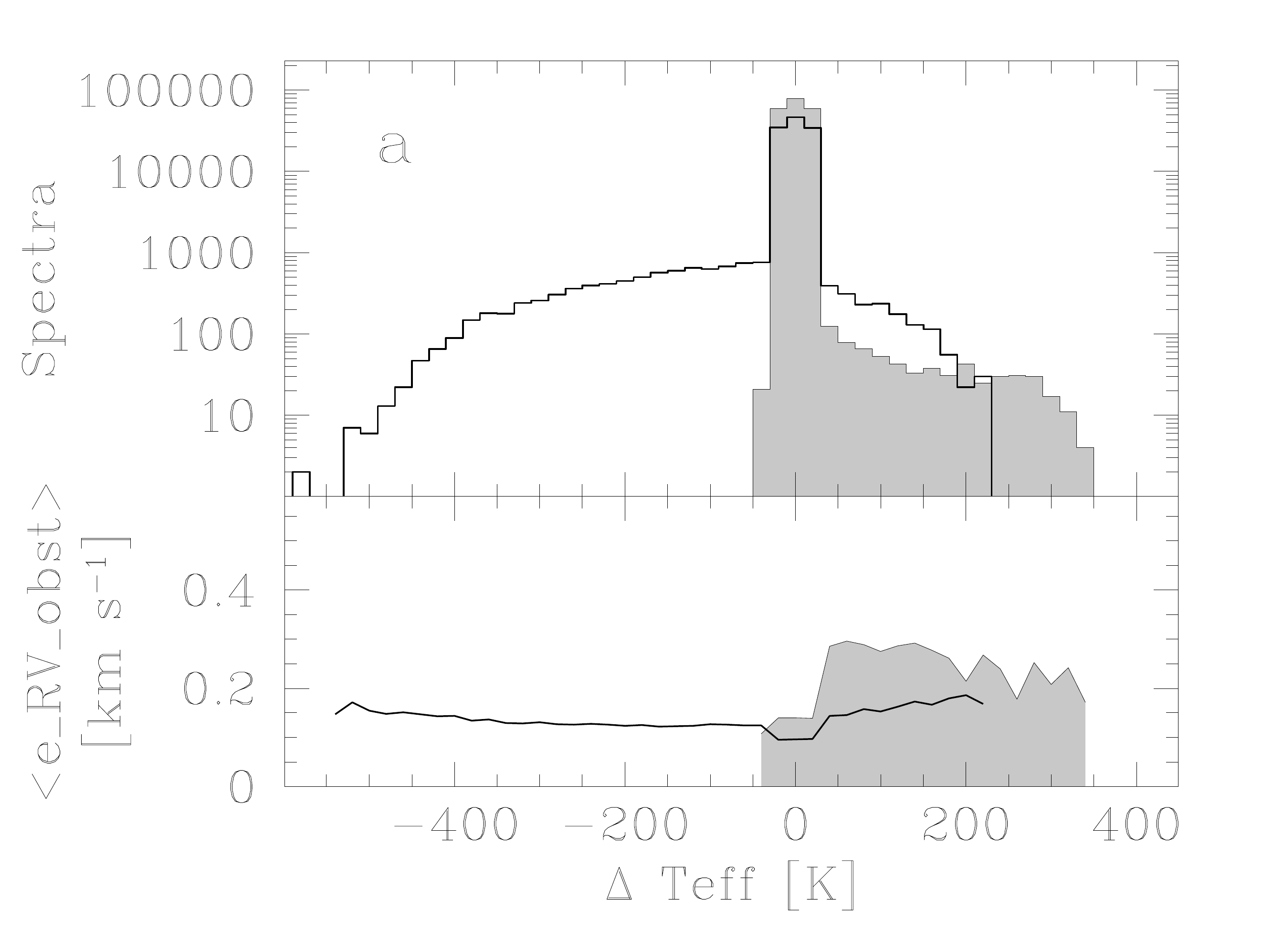}                   
  \includegraphics[width=0.68\columnwidth,height=1.0\columnwidth,angle=270]{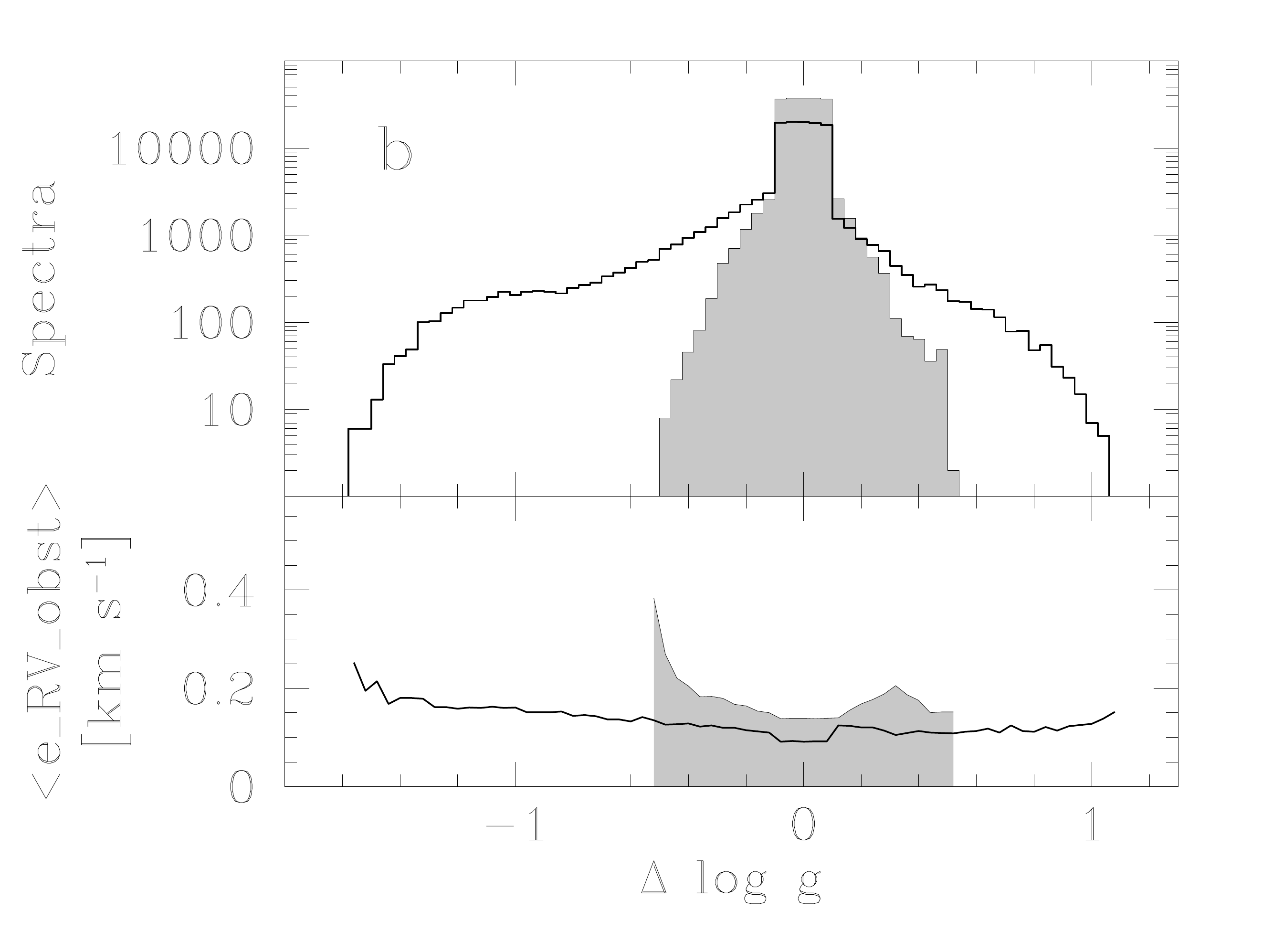}                   %
  \includegraphics[width=0.68\columnwidth,height=1.0\columnwidth,angle=270]{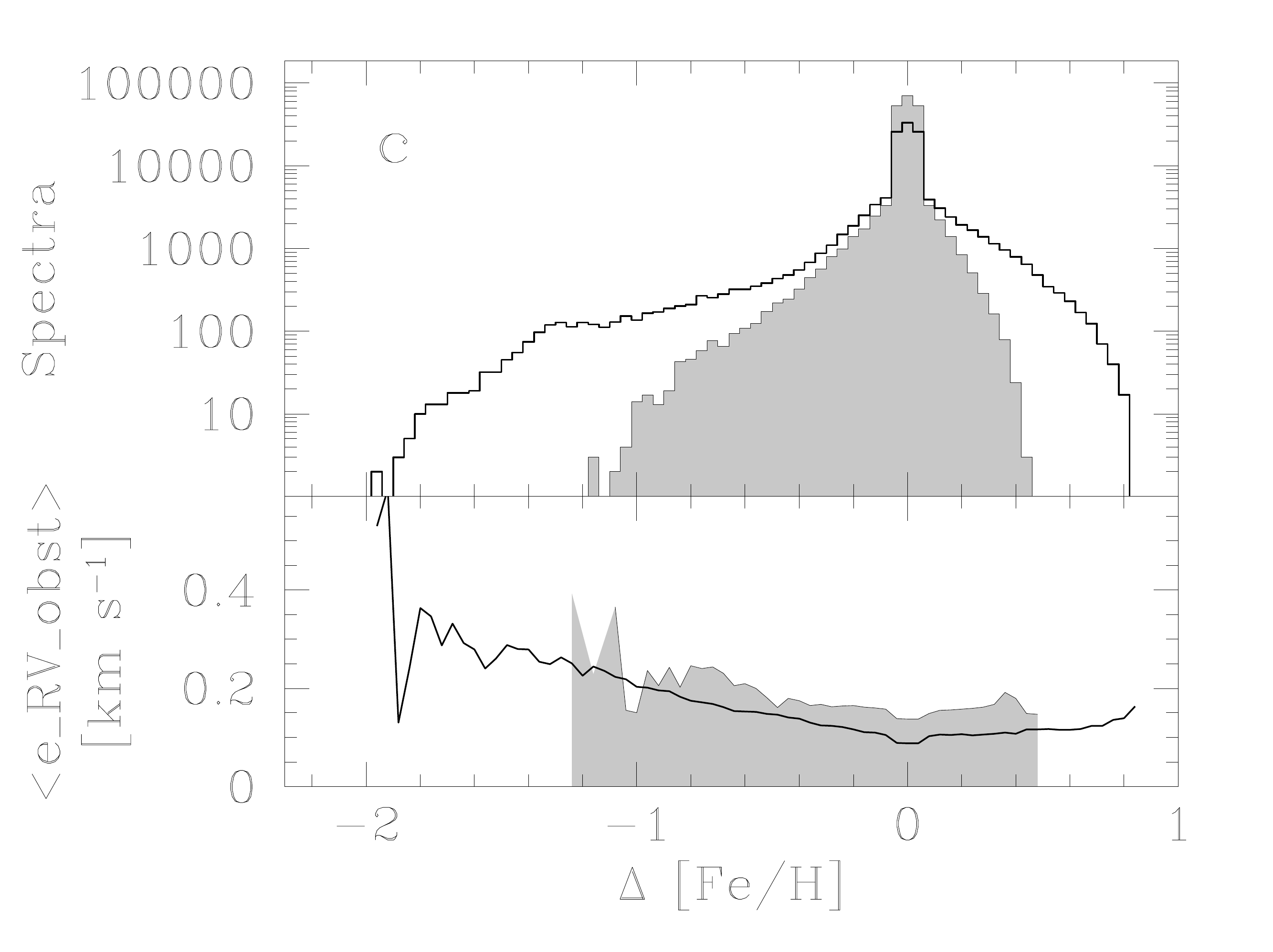}                   %
  \caption{Number of spectra (top) and their average final radial velocity errors (bottom, including the gravitational redshift error) as a function of difference between the Cannon parameter value and the one of the median spectrum used to calculate its radial velocity. Grey shading are plots for dwarfs and black lines are for giants, with Eq.\ 1 used to distinguish between the two. Individual panels plot results as a function of difference in temperature (a), gravity (b), and metallicity (c).}
  \label{Figbindistance}
\end{figure}

As a first step, we calculate the RV of the observed spectrum by cross-correlating it with the corresponding observed median spectrum in the appropriate bin. This process may not be straightforward. If the observed spectrum belongs to a bin of stellar parameters that is not well populated, the median spectrum of this bin could be affected by small number statistics in terms of noise and influence of possible stellar outliers. In such case it is better to use a median spectrum from a nearby well populated bin, instead. But this other bin should not be too far in parameter space as we want to avoid correlation against a median of very different spectra. After extensive tests we decided to use only those median spectra that have at $\gtrsim 100$ observed spectra in their bins. There are all together  1181 spectral bins in our observed median spectra library, which were calculated as a median of between  100 and 1116 observed spectra. If an observed spectrum does not belong to such a well populated bin, we use a median from the nearest sufficiently populated bin, adopting a weighted Manhattan distance metric
\begin{equation}
d = | \Delta \FeH | / a_0 +  | \Delta \logg | / a_1 + | \Delta \Teff | / a_2
\end{equation}
where $\Delta$ labels differences in parameter values between the observed spectrum and the values pertaining to a given bin. Measurement of RVs strongly depends on changes in temperature, gravity changes are less important, while metallicity changes have a small influence on relative strength of spectral lines. So we adopt the following values of inverse weights:  $a_0 = 0.1$~dex, $a_1 = 0.01$~dex, and $a_2 = 2$~K. Figure \ref{Figbindistance} shows that the differences between spectral parameters of the observed spectrum and those of a  median bin that was used for calculation of its radial velocity are still very small.  73\%\ of all spectra without warning flags belong to well populated median bins, and another 12\%\ of spectra have a distance $d<10$.

Radial velocity between observed spectrum and its median observed counterpart is measured in 20 wavelength ranges (Table \ref{Tableintervals}). This allows for rejection of outliers due to reduction problems or intrinsic peculiarities of a given observed spectrum, such as chromospheric activity. Wavelength intervals cover entire wavelength spans of each individual arm, except for the IR arm where a region polluted with strong telluric absorptions is avoided. Adjacent intervals have small overlaps in wavelength, but this does not give any extra weight to these overlapping regions because apodisation is used when calculating the correlation function. RVs in individual intervals are determined with the {\em fxcor} routine of the \textsc{Iraf}'s {\em rv} package, normalising both continua with a single cubic spline with $\pm 2.5 \sigma$ rejection criteria, using 20\%\ apodisation at both edges, and without use of any Fourier filtering. The measured RV is taken as the centre of a Gaussian fit to the upper half of the correlation peak, and eRV as the fitting error. 

Next we need to join measurements of RV$_i$ (and their statistical errors eRV$_i$) in these 20 intervals into a single RV and its error eRV. We do this iteratively using the weights $w_{Ni}$ defined by 
\begin{equation}
w_{Ni} = \frac{[(\mathrm{eRV}_i)^2 + (\mathrm{d}_N\mathrm{RV}_i)^2]^{-1}}{\sum_{i=1}^{20}[(\mathrm{eRV}_i)^2 + (\mathrm{d}_N\mathrm{RV}_i)^2]^{-1}}
\end{equation}
where $\mathrm{d}_N\mathrm{RV}_i$ is the difference between a current value of RV$_{N-1}$ in the $(N-1)$-th iteration and RV$_i$. The current value of RV$_N$ and its error eRV$_N$ are calculated as 
\begin{equation}
\mathrm{RV}_N = \sum_{i=1}^{20} w_{Ni} \mathrm{RV}_i
\end{equation}
and
\begin{equation}
\mathrm{eRV}_N^{2} = \sum_{i=1}^{20} w_{Ni}^2 [(\mathrm{eRV}_i)^2 + (\mathrm{d}_N\mathrm{RV}_i)^2] 
\end{equation}
The final estimate of RV and its error eRV is reached after  $N = 30$ iterations. This scheme is useful as it takes into account both the statistical errors of the RV measurements in individual intervals and their distance from the final averaged radial velocity. Table \ref{Tableintervals} reports averages of the final weights ($\overline{w_i}$) assigned to any of the 20 intervals, where averaging is done over  323,949 spectra in our database with zero values of reduction, synt, and parameter flags. For tabulation purposes, weights were rescaled so that their average value is 1.0. The table shows that different intervals do not contribute with the same weight to the final velocity. In particular, intervals of H$\alpha$  (3b) and the ones in the blue and green arm have higher weights, while some intervals located at the edges of the wavelength range of particular arms, and especially the ones in the infra-red arm, have significantly lower weights. The interval labelled 4e is affected by telluric absorptions which may be difficult to remove and which decrease the number of collected photons from a star. This explains the very low weight of this interval. Note that weights are determined for each spectrum, so Table \ref{Tableintervals} reports just average values. 

\begin{table*}
\caption{Wavelengths of intervals to measure radial velocity, their labels (lbl) and average relative weights ($\overline{w}$).}
\label{Tableintervals}
\begin{tabular}{lrccrccrccrccrcc}
arm &\multicolumn{3}{c}{interval}&\multicolumn{3}{c}{interval}&\multicolumn{3}{c}{interval}&\multicolumn{3}{c}{interval}&\multicolumn{3}{c}{interval}\\
 &lbl&$\mathrm{\lambda}$ [\AA ]&$\overline{w}$&lbl&$\mathrm{\lambda}$ [\AA ]&$\overline{w}$&lbl&$\mathrm{\lambda}$ [\AA ]&$\overline{w}$&lbl&$\mathrm{\lambda}$ [\AA ]&$\overline{w}$&lbl&$\mathrm{\lambda}$ [\AA ]&$\overline{w}$\\ \hline 
blue    &1a&4716--4756&1.29&1b&4751--4791&1.67&1c&4786--4826&1.20&1d&4821--4861&1.30&1e&4856--4896&1.38\\
green &2a&5655--5701&1.08&2b&5696--5742&1.30&2c&5737--5783&1.36&2d&5778--5824&0.94&2e&5819--5865&0.43\\
red     &3a&6480--6535&0.85&3b&6530--6585&2.19&3c&6580--6635&0.91&3d&6630--6685&0.81&3e&6680--6735&0.69\\
IR       &4a&7696-7737&0.62&4b&7732-7773&0.64&4c&7768-7809&0.78&4d&7804-7845&0.41&4e&7840-7881&0.14\\
                                                                                     \hline
\end{tabular}
\end{table*}

\subsection{Radial velocity of the observed median spectrum}

Although the observed median spectra are obtained as a combination of many observed individual spectra, there is still no guarantee that they have a net zero RV. The matter is more complicated, because most of the observed stars have convective atmospheres with macroscopic upstream and downstream motions \citep{gray92}. Even for a star at rest this implies that a line formed in the upwelling is blueshifted, while the one formed in the sinking gas between the convective upswells is redshifted. Effects of upstream and downstream motions do not cancel out, so lines emerging from the moving atmosphere are not centred around zero velocity, also their profile bisectors can be inclined \citep{allende98,allende13,asplund00}. Our spectra have a limited resolving power of $R=28,000$, but a very high S/N ratio for the median observed spectra. The goal of high accuracy in the derived RVs does not allow for a simplistic assumption that spectral lines are symmetrical and centred on the velocity of the object's barycentre. 

To address this concern, we use a synthetic spectral library \citep{chiavassa18} which takes 3-dimensional convective motions into account. It has been computed using the radiative transfer code Optim3D \citep{chiavassa09} for the \textsc{Stagger}-grid, a grid of 3D radiative hydrodynamical simulations of stellar convection \citep{magic13}. Synthetic spectra have been calculated for the purpose of this study at an original resolving power of 60,000 and then degraded to $R=28,000$ using a Gaussian convolution. In the process we checked that observed medians actually have the best match to synthetic spectra at this resolving power (to within 10\%). Thus the median observed spectra are not smeared and basically retain the resolving power of the original observations.  

The synthetic grid is coarser than the one of median observed spectra. We therefore used a linear interpolation of computed spectra,  which consist of a Solar spectrum plus a grid with a step of 500~K in temperature and 0.5~dex in gravity, assuming Solar metallicity. This moderate spectral mismatch has little influence on the accuracy of the derived RV (the lower panels in Fig.~\ref{Figbindistance}a-c demonstrate this for individual observed spectra). The observed median spectra are virtually noiseless, so the results are better than what would be obtained by direct correlation of individual observed spectra with a synthetic library. 

Values of RV$_{med}$ (RVs of median observed spectra vs.\ the interpolated synthetic library) have been determined in the same way as explained above for individual observations.  Their formal errors eRV$_{med}$ are equal to $0.06 \pm 0.03$~\kms. Average values of the median velocities vs.\ the synthetic library are $\overline{\mathrm{RV}_{med}} = 0.45 \pm 0.11$~\kms\ for dwarfs and $\overline{\mathrm{RV}_{med}} = 0.36 \pm 0.18$~\kms\ for giants (see Fig.~\ref{FigRVmedRVgrav}a). These velocity shifts are discussed in Sec.~\ref{Sec55}.

\subsection{Gravitational shifts}

In the absence of any real motion along the line of sight, the spectra of stars would still appear to suggest that they are receding from the Earth. This apparent recession is the result of the gravitational redshifting of the star's light, which originates at a distance $R$ from centre of mass $M$. The corresponding gravitational velocity shift equals 
\begin{equation}
\mathrm{RV}_{grav} = GM/(Rc) 
\end{equation}
and reaches 0.636~\kms\ for a Solar-type star. This equation holds for an observer at infinity if other gravitational potentials are negligible. While the effects of blueshift due to the fact that Earth is located within the Solar system and differences in the general Galactic potential are small, additional components need to be considered if the star is a member of a strongly bound system, e.g.\ a massive globular cluster or a dwarf galaxy (where this shift can reveal information on the distribution of dark matter). 

The value of the $M/R$ ratio is generally not known \citep{pasquini11}. We used a complete set of PARSEC isochrones for scaled Solar metallicity \citep{bressan12} and values of spectroscopic parameters $\Teff$, $\logg$, and $\FeH$ to determine the value of the $M/R$ ratio for each observed spectrum and hence its gravitational redshift. Its error was calculated from the errors on the spectroscopic parameters, assuming that they cannot be smaller than 50~K in temperature, 0.1~dex in gravity and 0.05~dex in metallicity. In the near future, one could use the spectroscopically determined $\Teff$ and the value of luminosity $L$ inferred from the trigonometric parallax from Gaia DR2 to derive stellar radius even more accurately. To assist the user we therefore report RVs with and without gravitational redshift.

\subsection{Derivation of the final values}

The final value of RV, excluding the gravitational redshift, equals  
\begin{equation}
\mathrm{RV}\_\mathrm{nogr}\_\mathrm{obst} = \mathrm{RV}_N + \mathrm{RV}_{med}
\end{equation}
and that taking gravitational redshift into account is
\begin{equation}
\mathrm{RV}\_\mathrm{obst} = \mathrm{RV}\_\mathrm{nogr}\_\mathrm{obst} - \mathrm{RV}_{grav}
\end{equation}

\begin{figure}
  \includegraphics[width=\columnwidth,height=1.0\columnwidth,angle=0]{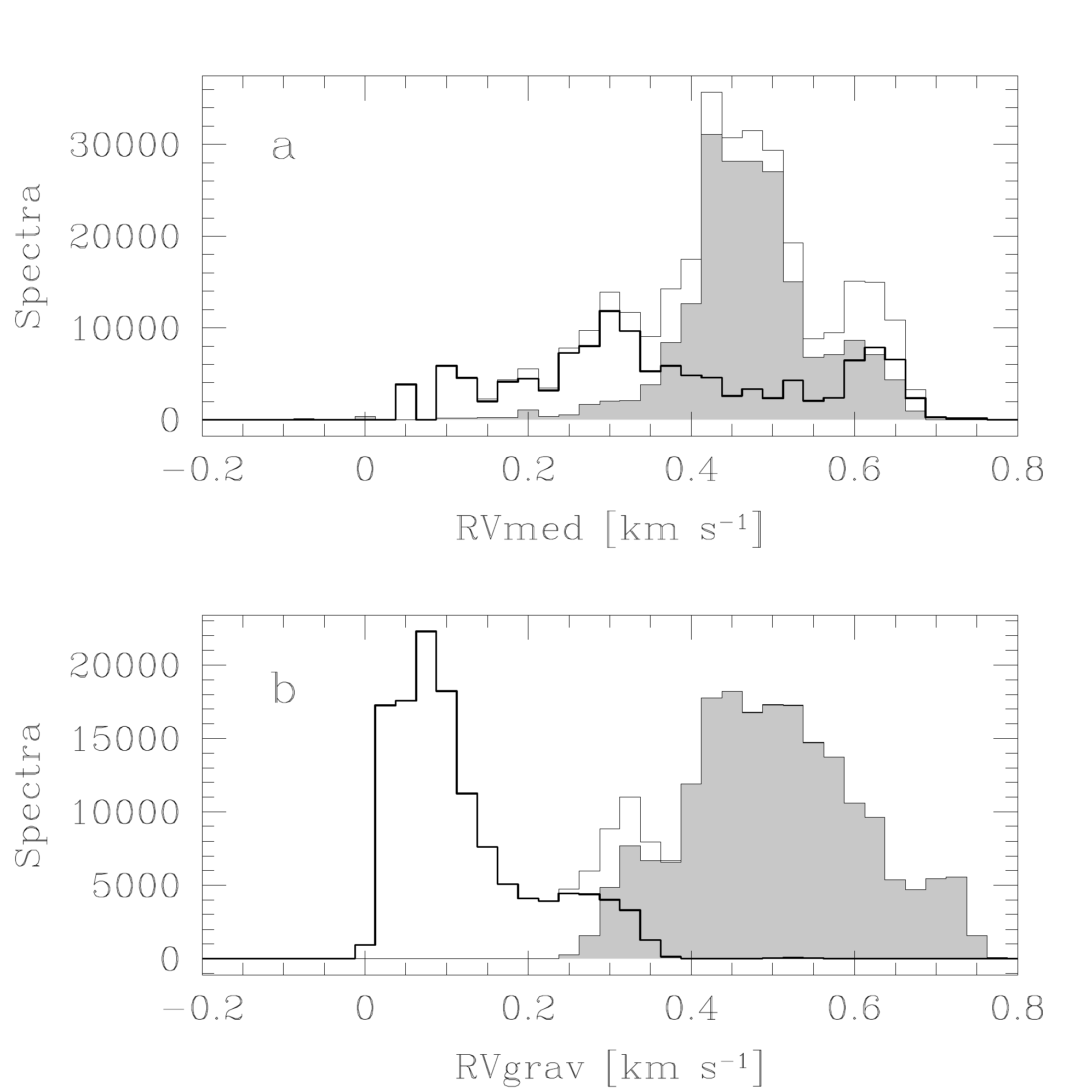} 
  \caption{ Distribution of median radial velocities (a) and gravitational redshifts (b) for spectra with zero values of all three warning flags. The three histograms are for dwarfs (filled grey), giants (thick lines), and their sum. The line dividing these two classes is defined in Eq.\ 1. Note that median radial velocities, which reflect convective blueshifts, and gravitational redshifts nearly cancel each other in dwarfs. }
  \label{FigRVmedRVgrav}
\end{figure}

\begin{figure}
  \includegraphics[width=0.68\columnwidth,height=1.0\columnwidth,angle=270]{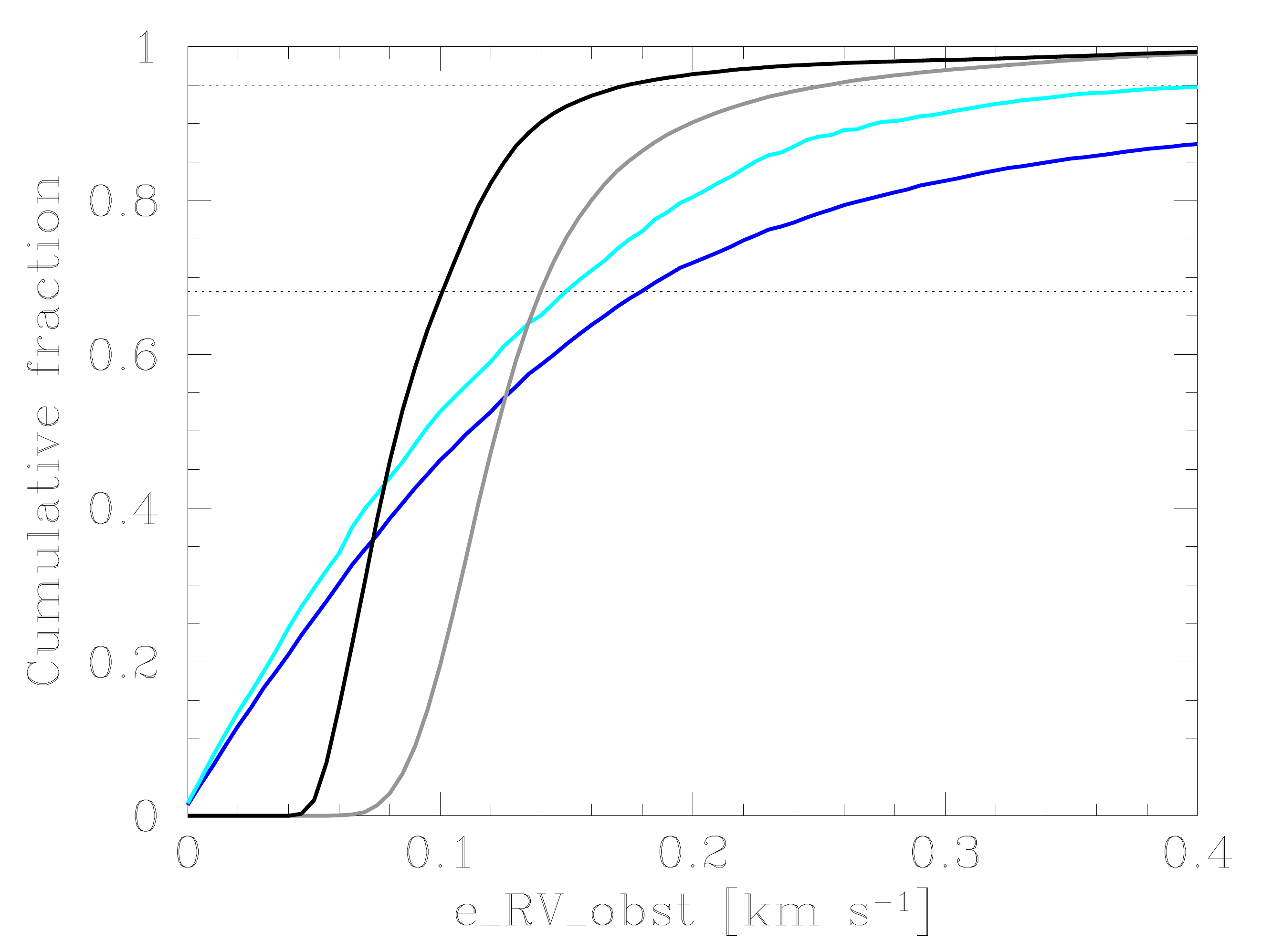} 
  \caption{Upper two curves: Cumulative histogram of formal radial velocity errors (e\_rv\_obst) for giants (black) and dwarfs (grey). Lower two curves: Cumulative distribution of standard deviation of actual repeated RV measurements of the same objects re-observed within a single night (cyan) or at any time-span (blue).  Horizontal lines mark 68.2\%\ and  95\%\ levels.}
  \label{Figcumulerrors}
\end{figure}

\label{Sec55}
Figure \ref{FigRVmedRVgrav} shows the distributions of $\mathrm{RV}_{med}$ and $\mathrm{RV}_{grav}$ for all spectra with zero values of reduction, synt, and parameter warning flags. Grey filled histograms are for dwarfs and the thick-lined ones are for giants. Ideally one would expect zero or very small values for $\mathrm{RV}_{med}$ (Fig.\ \ref{FigRVmedRVgrav}a), 
as median spectra were calculated after shifting observed spectra to approximate rest frame using the values of  RV\_synt (see Sec.\ \ref{Sec51}). But these velocities were computed by comparing observed spectra against AMBRE model spectra which do not allow for convective motions in stellar atmospheres, resulting in blueshifts of many spectral lines. As a consequence, velocities of median spectra of dwarfs are found to be positive when compared to 3D stellar atmosphere models of the \textsc{Stagger} grid \citep{chiavassa18}. The effect is less pronounced in giants and is related to the fact that only synthetic spectra of dwarfs were used in computation of RV\_synt values. 

Behaviour of gravitational redshifts (Fig.\ \ref{FigRVmedRVgrav}b) is as expected. Dwarf temperatures in the GALAH sample cluster around Solar values (Fig.\ \ref{Figparameterhistos}a) but their surface gravities are somewhat lower (Fig.\ \ref{Figparameterhistos}b), so the gravitational redshift for dwarfs is similar to or smaller than the Solar value (filled histogram of Fig.\ \ref{FigRVmedRVgrav}b). Large radius of giants makes their gravitational redshift much smaller (thick histogram of Fig.\ \ref{FigRVmedRVgrav}b). Note that histograms in Figs. \ref{FigRVmedRVgrav}a and \ref{FigRVmedRVgrav}b are similar, so effects of convective blueshifts and gravitational redshifts approximately cancel each other in dwarfs. But these two effects need to be taken into account if accuracy of radial velocities at a $\sim 0.1$~\kms\ level is to be achieved.

Formal uncertainties e\_RV\_obst and e\_RV\_nogr\_obst were calculated by Monte-Carlo propagation of individual errors. For the gravitational shift we propagated errors of stellar parameters, while any possible uncertainty of the isochrone calculations was neglected. Average values of the final uncertainty estimate (e\_RV\_obst) are shown in the lower panels of Fig.~\ref{Figbindistance}a-c. Cumulative histograms in Fig.~\ref{Figcumulerrors} demonstrate that typical uncertainties are around 0.1~\kms. Note that the uncertainty of RVs is somewhat larger in dwarfs than in giants because their gravitational shift is larger and also their spectra are not so rich. Error levels turn out to be similar to those expected from basic limitations on collected light: the photon noise precision from \citet{bouchy01} and \citet{beatty15} for S$/$N~$\sim 30$ per pixel, $\sim 10,000$ useful pixels and with a quality factor $1500 \lesssim Q \lesssim 3000$ is 0.03-0.06~\kms. 

The cyan line in Figure~\ref{Figcumulerrors} shows the cumulative distribution of standard deviations for  2613 stars observed twice during the same night, and the blue line shows  14,960 pairs observed at any time-span. Note that for two observations the standard deviation is half of the difference of measured RVs. If we assume that objects are not intrinsically variable within the same night one would expect that the cyan and black/grey curves would be identical. This is not the case for two reasons: (i) repeated observations within the same night were not planned in advance, so the observers frequently decided to carry them out because the first spectrum did not match the expected quality standards; (ii) fibres may not be illuminated in exactly the same way each time and incomplete scrambling then translates this variability into RV shifts. From Figure~\ref{Figcumulerrors} one may infer that the latter effect contributes at a level of 0.1~\kms, so that the sum of the formal RV error and additional errors due to any type of instability of the instrument is around 0.14~\kms. The difference between the cyan and blue curves is largely due to intrinsic RV variability of the observed stars, which may be a consequence of stellar multiplicity, pulsations, time-variable winds, or presence of spots.  

\section{Discussion}

\begin{figure}
 \begin{center}
  \includegraphics[width=0.86\columnwidth,height=1.06\columnwidth,angle=270]{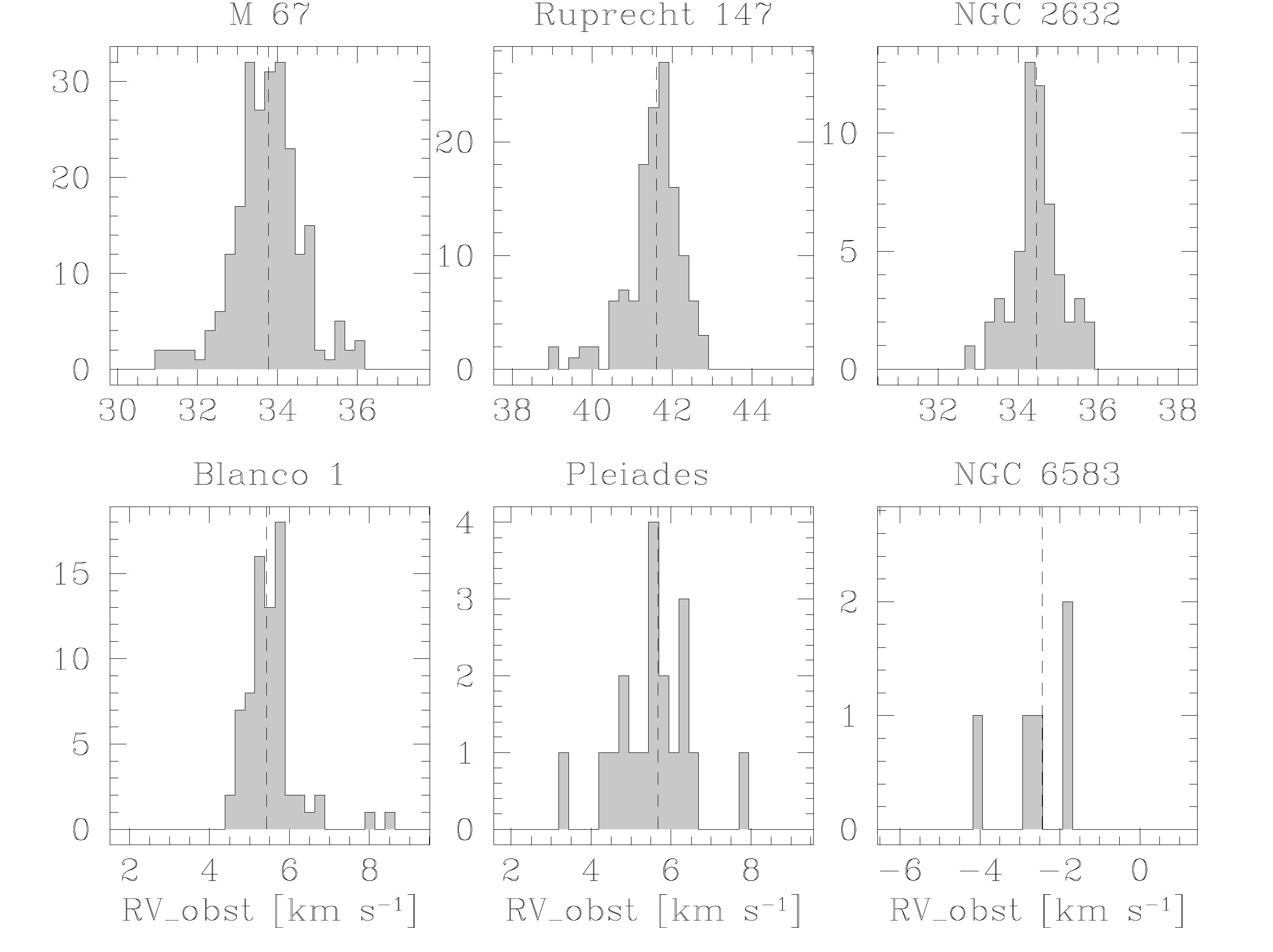}
 \end{center}
  \caption{Distribution of RVs for stars in clusters. Vertical dashed line marks the median RV.}
  \label{Figclusters}
\end{figure}

\begin{table}
\caption{Properties of stellar clusters. Table reports the number of spectra (N) of observed cluster member candidates, median observed radial velocity and metallicity, and standard deviations of all measurements around the median. The last column is RV from the literature (\citealt{kharchenko13,mermilliod09} or WEBDA database)].  All velocities are in \kms.}
\label{Tableclusters}
\begin{tabular}{lrccrcr}
            &              & Median &$\FeH$& Median& RV  & RV \\
cluster & N          & $\FeH$  &scatter &RV &scatter & liter.\\
 \hline
M 67      & 235 & $-0.02$&0.10& 33.77 & 1.09 & 33.6 \\       
Ruprecht 147 &  129 & $+0.08$&0.06&41.61 & 0.69 & 41.0\\
NGC 2632      & 56 & $+0.18$&0.08&34.47 & 0.62 & 33.4 \\
Blanco 1         & 73 & $+0.02$&0.13& 5.42 & 0.67 & 5.53 \\
Pleiades      & 19 & $-0.04$&0.10&5.66 & 0.97 & 5.67  \\
NGC 6583    & 5    & $+0.32$&0.08&-2.44  & 0.94 & 3.0 \\ 
                                                                                     \hline
\end{tabular}
\end{table}

\begin{figure}
  \includegraphics[width=1.05\columnwidth,height=1.2\columnwidth,angle=0]{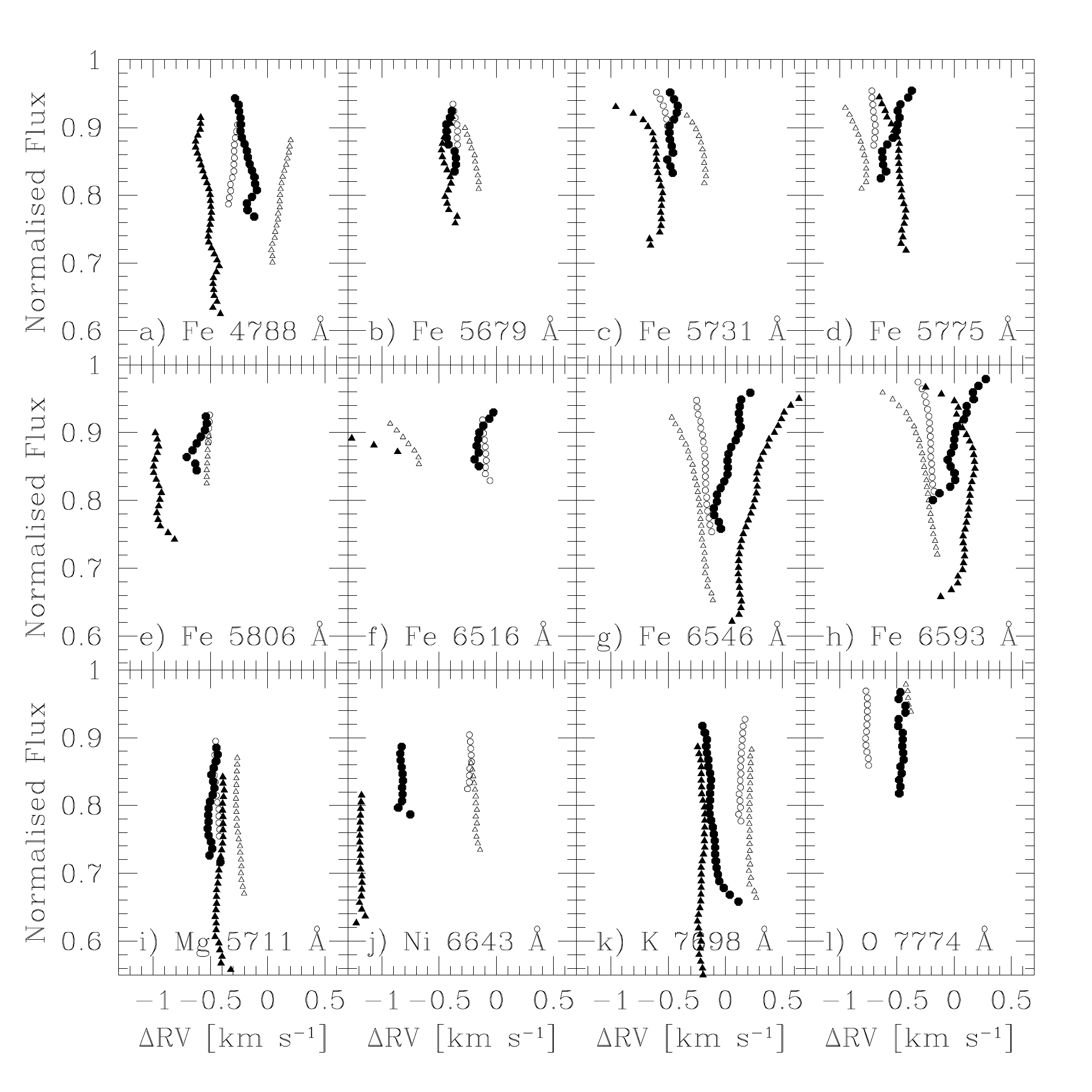} 
  \caption{Examples of observed and computed velocities of bisectors of iron lines for different levels of normalised continuum flux.
  Filled symbols are for observed median spectra and open ones for their synthetic counterparts. Circles label a dwarf star ($\Teff = 6000$~K, $\logg = 4.5$~dex), and triangles a  subgiant star ($\Teff = 4850$~K,  $\logg = 3.6$~dex), in both cases  with a Solar metallicity.}
  \label{Figbisectors}
\end{figure}

In the previous Section we showed that derived RVs have relatively small formal uncertainties and that they are repeatable at a level of $\sim 0.14$~\kms. Here we first check on velocity of stars that are members of known stellar clusters. Figure~\ref{Figclusters} shows the distribution of RVs for likely members which are located within the $r_2$ radius of \citet{kharchenko13}, with compatible proper motions from \textsc{Ucac-5} \citep{zacharias17} (plus one star with $\FeH = -0.63$ was removed from the  NGC~2632 list). The vertical line is the median RV, which is also reported in Table~\ref{Tableclusters}. Note that the scatter of RVs of cluster stars around the median is usually smaller than 1~\kms. Derived medians are  a refinement of values from the literature, with the exception of NGC~6583 which has an RV different from the published value \citep[][which is however based on only two stars]{kharchenko13} .   

\begin{figure*}
  \includegraphics[height=2.4\columnwidth,angle=0]{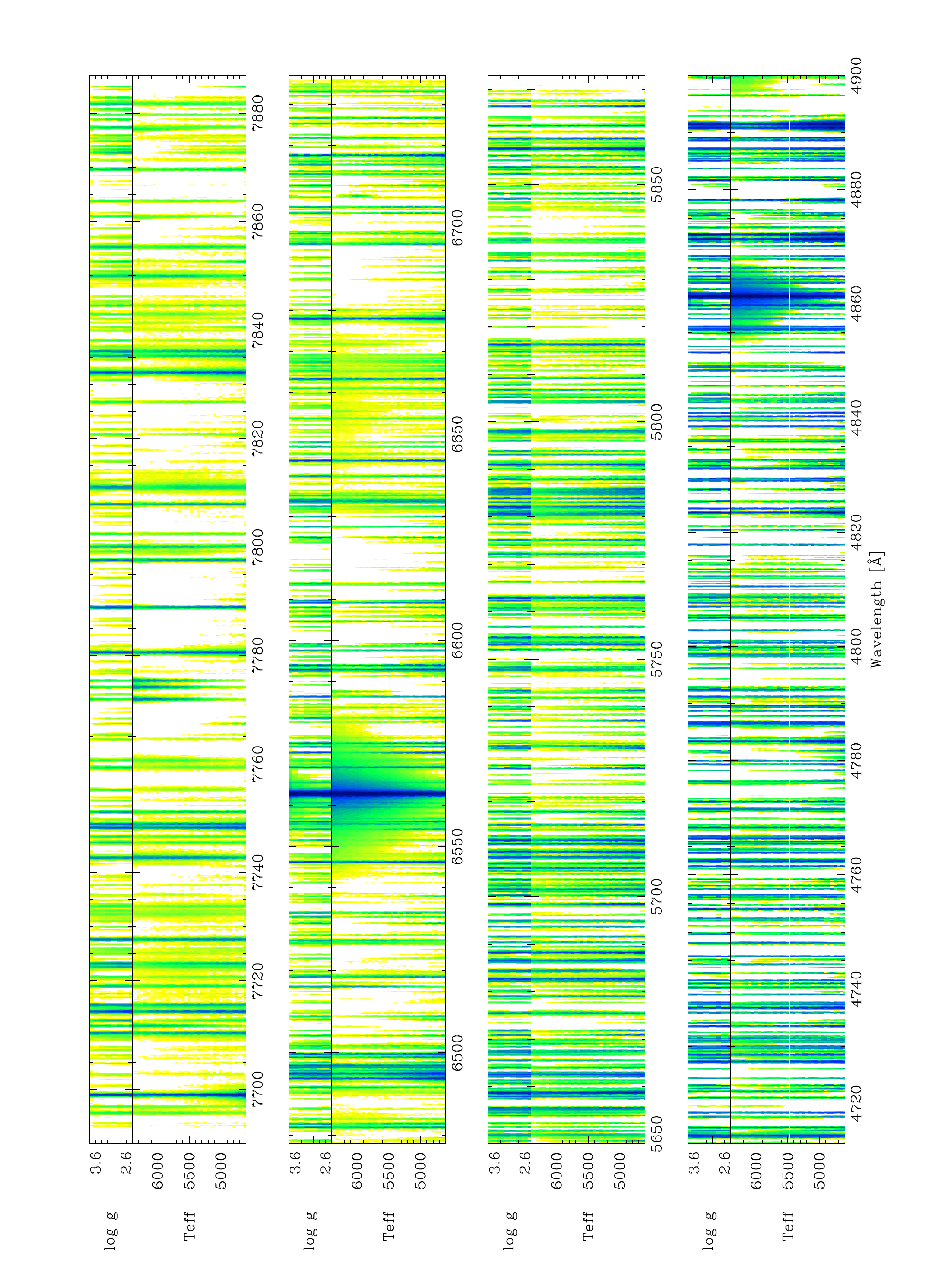}
  \caption{Cross-sections  through  the  multidimensional data-cube of observed median spectra for the four spectrograph arms. The bottom part of each panel shows a sequence of main-sequence spectra with different effective temperatures, indicated on the vertical axis. Similarly, the top part of each panel is a sequence of spectra along the red giant branch with indicated values of surface gravity. In both cases only spectra with Solar metallicity are plotted. Continuum regions are white, while absorption profiles are coloured with progressively darker colours, depending on their depth. }
  \label{Figimagemedians}
\end{figure*}

There are macroscopic motions in a convective stellar atmosphere, so absorption line profiles may be asymmetric. A detailed study of profile shapes is beyond the scope of this paper, but Fig.~\ref{Figbisectors} suggests that median observed spectra may be useful for this purpose. One of the ways to visualise velocity shifts is to compute line bisectors, i.e.\ mid-points between blue and red wing positions for a range of normalised flux levels. A symmetric line has a vertical bisector with a constant velocity, while asymmetries are characterised by inclined or curved bisectors \citep[see e.g.\ Fig.~6 in ][]{chiavassa18}. Note that the resolving power of GALAH spectra is $R=28,000$, so line asymmetries tend to be smeared. But not completely, as demonstrated by a number of inclined bisectors in Fig.~\ref{Figbisectors}. There we plot only lines which are, we believe, well isolated, as any line blends can produce inclined bisectors as well. In our experience blends tend to cause very strong asymmetries which is not the case for lines plotted in the Figure. As an example we plot bisectors for just two median observed spectra, a main sequence star a bit  hotter than the Sun and a  subgiant star, both with a Solar metallicity. These are compared to bisectors of their corresponding synthetic spectra \citep{chiavassa18}, downgraded to the same resolving power. All spectra are moved to the same reference frame, so that the cross-correlation function between the observed median spectrum and its synthetic counterpart has a peak at zero velocity ($\Delta$RV~$=0$). 

Fig.~\ref{Figbisectors} shows some nice matches between observed median spectra and their synthetic counterparts. Such are lines of Fe~I 5679, Fe~II 6516, and Mg~I 5711 for both Solar type and  subgiant stars, and  Fe~I 5731, and 5806  for Solar type dwarfs. In many other cases the shapes of observed and computed bisectors are nicely matching,  though both blue- and red-shifts are present. For the dozen lines shown in Fig.~\ref{Figbisectors} velocity difference between observed and synthetic spectra is $+0.0 \pm 0.2$~\kms\ for dwarfs and $-0.2 \pm 0.4$~\kms\ for subgiants.  In a few cases an offset may be due to different assumed central wavelengths; for observed medians we used the line list from \citet{buder18}.  A different continuum normalisation or presence of unknown line blends may be causing bisector shifts, but not with a constant offset: both of these offsets are very small close to line centres but get much larger as we increase flux levels toward line wings. We therefore suggest that offsets between synthetic and observed spectra are real and reflect stellar dynamical activity. 

\section{Data products}

The tabular results of this work are reported in four columns, which are published as part of the GALAH DR2 data release \citep{buder18}. These are the values of RV\_obst and RV\_nogr\_obst, and their errors e\_RV\_obst and e\_RV\_nogr\_obst, all in units of \kms.  The GALAH DR2 catalogue and documentation is available at \url{https://galah-survey.org} and at  \url{https://datacentral.aao.gov.au/docs/pages/galah/}, through a search form or ADQL query. The catalogue is also available through TAP via \url{https://datacentral.aao.gov.au/vo/tap}.  Some Python tools for processing the data are available online in an open-source repository\footnote{\url{https://github.com/svenbuder/GALAH_DR2}}.

An additional result of this work is a table of 1181 median observed spectra, also published on the GALAH website \url{https://galah-survey.org}. Their median velocity (RV$_{med}$) and gravitational redshifts are taken into account, so these spectra are at rest with respect to the stellar barycentre. Figure \ref{Figimagemedians} shows an illustrative cross-section across the data cube for stars with Solar metallicity. Median observed spectra  should be useful to compare results of any synthetic spectra library to an extensive set of observed spectra in the visible range. Similarly, the radial velocity of a star observed in the visible range at a similar resolving power to GALAH can be derived using the median observed spectral library as a reference. Note that the normalisation of these spectra is approximate, so the user is strongly encouraged to renormalise {\em both} spectral sets which are being compared with the same normalisation function and continuum fitting criteria before any comparison is made.  

\section{Conclusions}

Extensive observations of the GALAH spectroscopic stellar survey were used to derive and publicly release accurate values of radial velocities of  336,215 individual stars. Important effects of internal convective motions within a stellar atmosphere and of gravitational redshift of light as it leaves the star were taken into account.  Note that some objects may suffer from systematic errors due to observation problems or intrinsic peculiarity of the object, including binarity or presence of emission lines. Such problems are properly flagged. Our public release contains 212,734 objects which have all warning flags set to 0, so their radial velocities are the most reliable. 

Typical uncertainties  of radial velocities are at a 0.1~\kms\ level, with repeated observations of the same object consistent with an internal uncertainty of $\sim 0.14$~\kms. At a spectral resolving power of 28,000 this amounts to $1/77$ of the resolution element, hence close to what is achievable  for a random magnitude-limited sample of stars observed with a fibre-fed spectrograph  attached to a wide field-of-view telescope that is not particularly adapted for high precision radial velocity work\footnote{Note that \citet{loeillet08} achieved detection of RV variations with a treshold of $0.03$~\kms\ on 5 consecutive nights using  the FLAMES instrument on VLT which has a much narrower field of view and monitors instrument stability with simultaneous arc exposures. This precision can be compared to our formal errors on $RV_N$, which are $0.09 \pm 0.06$~\kms\ for dwarfs and $0.06 \pm 0.04$~\kms\ for giants. Mapping of spectrograph aberrations and wavelength solution using a photonic comb \citep{kos18} is expected to further improve the GALAH results.}. These error levels are important, as they would allow one to address internal dynamics of stellar clusters, associations, or streams, an important area of research, which is opening up with the advent of Gaia DR2. We use observed members of six open clusters to measure their radial velocities and show that the internal velocity dispersion of their members  { is $\sim 0.83 \pm 0.19$~\kms}. 

A separate result of this paper is a library of { 1181} observed median spectra with well defined values of stellar parameters. They have an extremely high S/N ratio and retain the initial resolving power of $R=28,000$. These spectra should be useful as an observed reference set to be compared to results of calculations of stellar models or for general radial velocity measurements in the visible range. Fig.~\ref{Figbisectors} suggests that they can be useful to study asymmetries of spectral line profiles. Both the radial velocities and the library of observed median spectra are published on the GALAH website, \url{https://www.galah-survey.org}. The catalog is available for querying at \url{https://datacentral.aao.gov.au}.

In future we plan to expand the number of radial velocities with new observations. We also plan to study median properties of spectra of peculiar stars in GALAH which have been identified with the tSNE spectra projection technique \citep{traven17}. They can improve our understanding of stellar peculiarities and of rapid and therefore rarely observed phases of stellar evolution. The magnitude limited GALAH survey, with its random selection function, high quality of recorded spectra and a large number of observed stars, is excellent for this purpose.

\section*{Acknowledgments}
{ We thank the referee for useful comments on the initial version of the manuscript.}
The GALAH survey is based on observations made at the Australian Astronomical Observatory, under programmes A/2013B/13, A/2014A/25, A/2015A/19, A/2017A/18. We acknowledge the traditional owners of the land on which the AAT stands, the Gamilaraay people, and pay our respects to elders past and present. TZ, GT, and K\v{C} acknowledge financial support of the Slovenian Research Agency (research core funding No.\ P1-0188 and project N1-0040). TZ acknowledges the grant from the distinguished visitor programme of the RSAA at the Australian National University. JK is supported by a Discovery Project grant from the Australian Research Council (DP150104667) awarded to J.\ Bland-Hawthorn and T.\ Bedding. ARC acknowledges support through the Australian Research Council through grant DP160100637. LD, KF and Y-ST are grateful for support from Australian Research Council grant DP160103747. SLM acknowledges support from the Australian Research Council through grant DE140100598. LC is the recipient of an ARC Future Fellowship (project number FT160100402). Parts of this research were conducted by the Australian Research Council Centre of Excellence for All Sky Astrophysics in 3 Dimensions (ASTRO 3D), through project number CE170100013.

\bigskip

\noindent
\hrulefill \\
$^{1}$Faculty of Mathematics and Physics, University of Ljubljana, Jadranska 19, 1000 Ljubljana, Slovenia\\
$^{2}$Sydney Institute for Astronomy, School of Physics, A28, The University of Sydney, NSW 2006, Australia\\
$^{3}$Universit\'{e} C\^{o}te d'Azur, Observatoire de la C\^{o}te d'Azur, CNRS, Lagrange, CS 34229, Nice, France\\
$^{4}$Max Planck Institute  for Astronomy (MPIA), Koenigstuhl 17, 69117 Heidelberg, Germany\\
$^{5}$Research School of Astronomy \& Astrophysics, Australian National University, ACT 2611, Australia\\
$^{6}$Center of Excellence for Astrophysics in Three Dimensions (ASTRO 3D), Australia\\
$^{7}$School of Physics and Astronomy, Monash University, Clayton 3800, Australia\\
$^{8}$ Faculty of Information Technology, Monash University, Clayton 3800, Victoria, Australia\\
$^{9}$Australian Astronomical Observatory, North Ryde, NSW 2133, Australia\\
$^{10}$Department of Physics and Astronomy, Uppsala University, Box 516, SE-751 20 Uppsala, Sweden\\
$^{11}$School of Physics, UNSW, Sydney, NSW 2052, Australia\\
$^{12}$INAF, Osservatorio Astronomico di Padova, Sede di Asiago, I-36012 Asiago (VI), Italy\\
$^{13}$Department of Physics and Astronomy, Macquarie University, Sydney, NSW 2109, Australia\\
$^{14}$Department of Astronomy, University of Virginia, Charlottesville, VA 22904-4325, USA\\
$^{15}$Stellar Astrophysics Centre, Department of Physics and Astronomy, Aarhus University, DK-8000, Aarhus C, Denmark\\
$^{16}$University of Southern Queensland, Computational Engineering and Science Research Centre, Toowoomba, Queensland 4350, Australia\\
$^{17}$ICRAR, The University of Western Australia, 35 Stirling Highway, Crawley, WA 6009, Australia\\
$^{18}$Center for Astrophysical Sciences and Department of Physics and Astronomy, The Johns Hopkins University, Baltimore, MD 21218, USA\\
$^{19}$Institute for Advanced Study, Princeton, NJ 08540, USA\\
$^{20}$Observatories of the Carnegie Institution of Washington, 813 Santa Barbara Street, Pasadena, CA 91101, USA\\
$^{21}$Department of Astrophysical Sciences, Princeton University, Princeton, NJ 08544, USA\\
$^{22}$Western Sydney University, Locked Bag 1797, Penrith, NSW 2751, Australia\\

\end{document}